\documentclass[12pt]{iopart}

\usepackage[english]{babel}
\usepackage{revsymb}
\usepackage{amsfonts}
\usepackage{amssymb}
\usepackage{graphicx}

\newcommand{\Pout}{P_{\mathrm{out}}}
\newcommand{\Wout}{W_{\mathrm{out}}}

\newcommand{\Qin}{Q_{\mathrm{in}}}
\newcommand{\hyp}{\mathrm{_{2\,}\!F\!_{1}}}

\newcommand{\tper}{t_{\mathrm{p}}}

\newcommand{\matice}[1]{\mathbb{#1}}
\newcommand{\J}{\matice{I}}
\newcommand{\Rp}{\matice{R}_+}
\newcommand{\Rm}{\matice{R}_-}
\newcommand{\Rc}{\matice{R}_\mathrm{p}}

\newcommand{\ket}[1]{\left| #1 \right\rangle}
\newcommand{\braket}[2]{\left\langle #1|#2\right\rangle}

\newcommand{\matr}[4]{\left(\begin{array}{cc} #1 & #2 \\ #3 & #4  \end{array}\right)} %4x4 matrix
\newcommand{\der}[1]{\mathrm{d}#1}

\newcommand{\RW}[1]{\textsf{#1}}

\newcommand{\prob}[1]{\mathop{\mathrm{Prob}}\, \left \{ #1 \right \}}
\newcommand{\Nder}[2]{\frac{\der{#1}}{\der{#2}}} %first derivative
\newcommand{\Pder}[2]{\frac{\partial{#1}}{\partial{#2}}} %first partial derivative
\newcommand{\JmJ}{\matr{1}{0}{0}{-1}}

\begin{document}

\title[]{Energetics and performance of a microscopic heat engine based
  on exact calculations of work and heat distributions}

\author{Petr Chvosta$^{1}$, Mario Einax$^{2}$, Viktor Holubec$^1$,
  Artem Ryabov$^1$ and Philipp Maass$^{2,3}$}

\address{$^{1}$ Department of Macromolecular Physics, Faculty of
  Mathematics and Physics, Charles University,
  V~Hole\v{s}ovi\v{c}k\'ach~2, CZ-180~00~Praha, Czech Republic}

\address{$^{2}$ Institut f\"ur Physik, Technische Universit\"at
  Ilmenau, 98684 Ilmenau, Germany}

\address{$^3$ Fachbereich Physik, Universit\"at Osnabr\"uck,
  Barbarastra\ss e 7, 49069 Osnabr\"uck, Germany}
\eads{\mailto{chvosta@kmf.troja.mff.cuni.cz},
  \mailto{philipp.maass@uni-osnabrueck.de}}

%\ead{philipp.maass@uni-osnabrueck.de}
%\ead{chvosta@kmf.troja.mff.cuni.cz}
%\homepage{}

\date{10 December 2009}

\begin{abstract}

  We investigate a microscopic motor based on an externally controlled
  two-level system. One cycle of the motor operation consists of two
  strokes. Within each stroke, the two-level system is in contact with
  a given thermal bath and its energy levels are driven with a
  constant rate. The time evolution of the occupation probabilities of
  the two states are controlled by one rate equation and represent the
  system's response with respect to the external driving. We give the
  exact solution of the rate equation for the limit cycle and discuss
  the emerging thermodynamics: the work done on the environment, the
  heat exchanged with the baths, the entropy production, the motor's
  efficiency, and the power output. Furthermore we introduce an
  augmented stochastic process which reflects, at a given time, both
  the occupation probabilities for the two states and the time spent
  in the individual states during the previous evolution. The exact
  calculation of the evolution operator for the augmented process
  allows us to discuss in detail the probability density for the
  performed work during the limit cycle. In the strongly irreversible
  regime, the density exhibits important qualitative differences with
  respect to the more common Gaussian shape in the regime of weak
  irreversibility.

\end{abstract}

\pacs{02.50.Ey, 05.40.-a, 05.70.Ln}
%\pacs{02.50.Ey, 05.10.Gg, 05.40.-a, 05.40.Jc, 05.45.-a, 82.20.Mj}
%02.50.Ey    Stochastic processes
%05.10.Gg    Stochastic analysis methods (Fokker-Planck, Langevin,etc.)
%05.40.Jc    Brownian motion
%05.45.-a    Nonlinear dynamics and chaos (see also section 45
    %Classical mechanics of discrete systems; for chaos in fluid
    %dynamics, see 47.52.+j; for chaos in superconductivity, see
    %74.40.De)
%05.70.Ln Nonequilibrium and irreversible thermodynamics
%         chemistry, see 82.65.+r, for surface magnetism, see 75.70.Rf)
%05.10.-a Computational methods in statistical physics
%05.10.Ln Monte Carlo methods
%05.40.-a Fluctuation phenomena, random processes, noise, Brownian motion
%82.20.Wt Computational modeling: simulation (Physical chemistry and chemical
%physics)
%82.20.Mj does not exist anymore

%\submitto{\JSTAT}

\maketitle

\section{Introduction}
\label{sec:I} Non-equilibrium phenomena in the presence of
time-varying external fields are of vital interest in many areas of
current research \cite{Evans/Searles:2002,Ritort:2008,Seifert:2008}.
Examples are aging and rejuvenation effects in the rheology of
soft-matter systems and in the dynamics of spin glasses, relaxation
and transport processes in biological systems such as molecular
motors, ion diffusion through membranes, or stretching of DNA
molecules, driven diffusion systems with time-dependent bias, and
nano-engines. With minimization of the system size thermal
fluctuations become increasingly relevant. In these systems it is
useful to introduce microscopic heat and work quantities as random
variables whose averages yield the common thermodynamic quantities.
Averages over functions of these microscopic heat and work quantities
yield generalized fluctuation theorems
\cite{Bochkov/Kuzovlev:1981,Evans/Cohen/Moriss:1993,Gallavotti/Cohen:1995,Jarzynski:1997,Crooks:1999,Maes:2003,
  Hatano/Sasa:2001,Speck/Seifert:2004,Seifert:2005,Schuler/etal:2005,Esposito/Mukamel:2006}.
In this context mesoscopic engines operating between different heat
baths under non-equilibrium conditions have received increasing
attention. The variety of models can be roughly classified according
to the dynamical laws involved. In the case of the classical
stochastic heat engines, the state space can either be discrete or
continuous (c.f., for example,
\cite{Sekimoto/etal:2000,Broeck/etal:2004,Schmiedl/Seifert:2008} and
the references therein). Examples of the quantum heat engines are
studied, e.g., in \cite{Henrich/etal:2007, Allahverdyan/etal:2008}.

The traditional consideration of efficiency of heat engines operating
between two baths at temperatures $T_{1}$ and $T_{2}$ leads to the
Carnot upper bound $\eta_{\rm C}= 1-T_1/T_2$. The bound is only
achieved under reversible conditions where the state changes require
infinite time and hence the power output is zero. Real heat engines
generate a finite power output
$P_{\mathrm{out}}=W_{\mathrm{out}}/t_{{\rm cycle}}$, i.e., they
perform work $W_{\mathrm{out}}$ during a cycle of a finite duration
$t_{{\rm cycle}}$. Thus an appropriate way to characterize the engines
is to compare their efficiencies at maximum power. On the macroscopic
level this quantity is roughly bounded by the Curzon-Ahlborn value
$\eta_{\rm CA} =1 - \sqrt{T_1/T_2}$ \cite{Curzon-Ahlborn:1975}.
Alternative expressions for quantifying efficiency have been discussed
\cite{Schmiedl/Seifert:2008} which are based on \emph{mean}
quantities, e.g., on the mean work done during the operational cycle.
On the mesoscopic level, the work is inherently a fluctuating quantity
and one should be able to calculate not only its mean value but also
its fluctuation properties.

In this paper we study a simple model of mesoscopic heat engines
operating between two different heat baths under non-equilibrium
conditions. The working medium consist of a two-level system. The
cycle of operation includes just two isothermal branches, or strokes.
Within each stroke, the system is driven by changing the energies of
the two states and we assume a constant driving rate, i.e.\ a linear
time-dependence of the energies. The response of the working medium is
governed by a master equation with time-dependent transition rates.
The specific form of the rates guarantees that, provided the two
energies were fixed, the system would relax towards the Gibbs
equilibrium state. Of course, during the motor operation, the Gibbs
equilibrium is never achieved because the energies are cyclically
modulated. At a given instant, the system's dynamics just reflects the
instantaneous position of the energy levels. After a transient regime,
the engine dynamics approaches \emph{limit cycle} with the periodicity
of the driving force. We will focus on the properties of this limit
cycle. In particular, we calculate the distribution of the work during
the limit cycle.

Our two-isotherm setting imposes one important feature which is worth
emphasizing. As stated above, at the end of each branch we remove the
present bath and we allow the thermal interaction with another
reservoir. This exchange of reservoirs necessarily implies a finite
difference between the new reservoir temperature and the actual system
(effective) temperature. Even if the driving period tends to infinity,
we shall observe a positive entropy production originating from the
relaxation processes initiated by the abrupt change of the contact
temperature. Differently speaking, our engine operates in an
inherently irreversible way and there exists no reversible limit.

The paper is organized as follows. In section~\ref{sec:II} we solve
the dynamical equation for the externally driven working medium. For
the sake of clarity we first give the solution just for an
unrestricted linear driving protocol using a generic driving rate and
a generic reservoir (section~\ref{subsec:IIa}). Thereupon, in
section~\ref{subsec:IIb}, we particularize the generic solution to
individual branches and, using the Chapman-Kolmogorov condition, we
derive the solution for the limit cycle. In section~\ref{sec:III} we
employ the recently derived \cite{Chvosta/etal:2007} analytical result
for the work probability density under linear driving. Again, we first
give the result for the generic linear driving and then we combine two
such particular solutions into the final work distribution valid for
the limit cycle. The results from section~\ref{sec:II} and
section~\ref{sec:III} enable a detailed calculation of the energy and
entropy flows during the limit cycle in section~\ref{sec:IV} and allow
for a discussion of the engine performance in section~\ref{sec:V}.

\section{Description of the engine and its limit cycle}
\label{sec:II} Consider a two-level system with time-dependent
energies $E_{i}(t)$, $i=1,2$, in contact with a single thermal
reservoir at temperature $T$. In general, the heat reservoir
temperature $T$ may also be time-dependent. The time evolution of the
occupation probabilities $p_{i}(t)$, $i=1,2$, is governed by the
master equation \cite{Gammaitoni/etal:1998} with time-dependent
transition rates specified by the reservoir temperature and by the
external parameters. To be specific the dynamics of the system is
described by the time inhomogeneous Markov process $\mathsf{D}(t)$
assuming the value $i$, $i=1,2$, if the system resides at time $t$ in
the $i$th state. Explicitly, the master equation reads
\begin{equation}
\label{pauliequation} \frac{\mathrm{d} }{\mathrm{d}
t}\mathbb{R}(t\,|\,t')=-
\left( \begin{array}{rr}
\lambda_{1}(t)&-\lambda_{2}(t)\\
-\lambda_{1}(t)& \lambda_{2}(t)
\end{array} \right)
\,\mathbb{R}(t\,|\,t')\,\,,\qquad\mathbb{R}(t'\,|\,t')=\mathbb{I}
\, ,
\end{equation}
where $\mathbb{I}$ is the unity matrix and ${\mathbb R}(t,t')$
the transition matrix with elements
$R_{ij}(t\,|\,t')=\langle\,i\,|\,{\mathbb
  R}(t\,|\,t')\,|\,j\,\rangle$, $i,j=1,2$. These elements are the
conditional probabilities
\begin{equation}
\label{rijdef} R_{ij}(t\,|\,t')={\rm
Prob}\left\{\,\mathsf{D}(t)=i\,|\,\mathsf{D}(t')=j\,\right\}\, .
\end{equation}
If we denote by $\phi(t')$ the initial state at time $t'$ with the
occupation probabilities $p_{i}(t')=\braket{\,i\,}{\phi(t')}$, the
occupation probabilities at the observation time $t$ are described by
the column vector
$\ket{p(t,t')}=\mathbb{R}(t\,|\,t')\,\ket{\phi(t')}$.

Due to the conservation of the total probability the system
(\ref{pauliequation}) can be reduced to just one non-homogeneous
linear differential equation of the first order. Therefore the master
equation (\ref{pauliequation}) is exactly solvable for arbitrary
functions $\lambda_{1}(t)$, $\lambda_{2}(t)$. The rates are typically
a combination of an attempt frequency to exchange the state multiplied
by an acceptance probability. We shall adopt the Glauber form
\begin{eqnarray} \fl
\label{transferrates}
\lambda_{1}(t)=\frac{\nu}{1+\exp\left\{-\beta(t)\left[E_{1}(t)-E_{2}(t)\right]\right\}}
\, , \;
\lambda_{2}(t)=\lambda_1(t)\exp\left\{-\beta(t)\left[E_{1}(t)-E_{2}(t)\right]\right\},
\end{eqnarray}
%\begin{equation}
%\label{transferrates}
%\lambda_{1}(t)=\nu\frac{1}{1+\exp\left\{-\beta(t)\left[E_{1}(t)-E_{2}(t)\right]\right\}}
%\,\,\,\,,\,\,\,\,
%\lambda_{2}(t)=\lambda_1(t)\exp\left\{-\beta(t)\left[E_{1}(t)-E_{2}(t)\right]\right\}
%\, ,
%\end{equation}
where $\nu^{-1}$ sets the elementary time scale, and
$\beta(t)=1/k_{B}T(t)$. The rates in equation~(\ref{pauliequation})
satisfy the (time local) detailed balance condition.

The general solution of the master equation (\ref{pauliequation}) for
the transfer rates (\ref{transferrates}) reads
\begin{eqnarray}
\fl \label{PMEsolution}
\mathbb{R}(t\,|\,t')=\mathbb{I}-\frac{1}{2} \left(
\begin{array}{rr}
1&-1\\
-1&1
\end{array} \right)
\left\{1-\exp\left[-\nu(t-t')\right]\right\}+ \frac{1}{2} \left(
\begin{array}{rr}
-1&-1\\
1&1
\end{array} \right) \,\xi(t,t')\, ,
\end{eqnarray}
where
\begin{equation}
\label{xidef} \xi(t,t')=\nu \int_{t'}^{t}\,\mathrm{d}
\tau\,\exp\left[-\nu(t-\tau)\right]\,
\tanh\left\{\frac{\beta(\tau)}{2}\,\left[E_{1}(\tau)-E_{2}(\tau)\right]\right\}
\, .
\end{equation}
The resulting propagator satisfies the Chapman-Kolmogorov condition
\begin{equation}
\label{Rchapmankolmogorov}
\mathbb{R}(t\,|\,t')=\mathbb{R}(t\,|\,t'')\mathbb{R}(t''\,|\,t')
\end{equation}
for any intermediate time $t''$. Its validity can be easily checked by
direct matrix multiplication. The condition simply states that the
initial state for the evolution in the time interval $[t'',t]$ can be
taken as the final state reached in the interval $[t',t'']$. This is
true even if the parameters of the process in the second interval
differ from those in the first one. Of course, if this is the case, we
should use an appropriate notation which distinguishes the two
corresponding propagators. This procedure will be actually implemented
in the paper. Keeping in mind this possibility, we shall first analyze
the propagator for a \emph{generic} linear driving protocol.

\subsection{Generic case -- linear driving protocol}
\label{subsec:IIa} Let us consider the linear driving protocol
$E_1(t)=h+v(t-t')$, and $E_2(t) = - E_1(t)$, where $h=E_{1}(0)$
denotes the energy of the first level at the initial time $t'$, and
$v$ is the driving velocity (energy change per time). The rates
(\ref{transferrates}) can then be written in the form
\begin{equation}
\label{transferrateslinear} \lambda_{1}(t)=\nu\frac{1}{1+c
\exp[-\Omega(t-t')]} \, , \,\,\,\, \lambda_{2}(t)=\nu\frac{c
\exp[-\Omega(t-t')]}{1+c \exp[-\Omega(t-t')]} \, ,
\end{equation}
where $\Omega=2\beta|v|$ is the temperature-reduced driving velocity,
and $c=\exp(-2\beta h |v|/v)$ incorporates the initial values of the
energies.

Under this linear driving protocol one can evaluate the definite
integral in (\ref{PMEsolution}) explicitly and rewrite the propagator
as
\begin{eqnarray}
\fl \label{PauliSolutionHeatBathGenericA}
\mathbb{R}(t\,|\,t')=\J-\matr{\phantom{-}1}{\phantom{-}0}{-
1}{\phantom{-}0} \left\{1 - \exp{\left [- \nu (t - t')\right ]}
\right \}+\matr{\phantom{-}1}{\phantom{-}1}{- 1}{- 1}
\gamma(t,t')\, ,
\end{eqnarray}
where
\begin{eqnarray}
\label{PauliSolutionHeatBathGenericAIntegral} \gamma(t,t') &=&
\nu\, c \int_{t'}^{t} \,\mathrm{d}\tau\, \exp{\left[- \nu (t -
\tau)\right]} \frac{\exp{\left(-\Omega
\tau\right)}}{1+ c \exp{\left(-\Omega \tau\right)}} \nonumber \\
&=& a\, c \exp{\left(- \nu t\right)} \int_{\Omega t'}^{\Omega t}
\mathrm{d}\tau\, \frac{\exp{\left[(a - 1) \tau \right]}}{1+ c
\exp{\left(- \tau \right)}}\, .
\end{eqnarray}
Here we have introduced the dimensionless ratio $a=\nu/\Omega$ of the
attempt frequency characterizing the time scale of the system's
dynamics and the time scale of the external driving, respectively.
Naturally, this ratio will describe the degree of irreversibility of
the process. Depending on the value $a\in(0,\infty)$, the explicit
form of the function $\gamma(t,t')$ reads
\begin{eqnarray}
\fl \label{gammaexact} \gamma(t,t')=\left\{\begin{array}{ll}
\frac{a c}{1-a}\exp{\left(-\nu t\right)} \left\{
\exp{\left[(a-1)\Omega
t'\right]}\,\hyp(1,1-a;2-a;-c\exp{\left(-\Omega t'\right)})
\right. &
\\[1ex]
\left. \hspace*{0.2cm}-\exp{\left[(a-1)\Omega
t\right]}\,\hyp(1,1-a;2-a;-c\exp{\left(-\Omega
t\right)}) \right\}\, , & a\in(0,1)\, , \\[2ex]
c\exp{\left(-\Omega
t\right)}\left[\Omega(t-t')+\ln{\displaystyle\frac{1+c\exp{\left(-\Omega
 t\right)}}{1+c\exp{\left(-\Omega t'\right)}}}\right ]\, , & a=1\,
 ,\\[2ex]
 \hspace*{0.2cm} \hyp(1,a;1+a;- \frac{1}{c} \exp{\left(\Omega t
   \right)}) 
& \\[1ex]
 -\exp{\left[-a\Omega(t-t')\right]}\,
\hyp(1,a;1+a;-\frac{1}{c}\exp{\left(\Omega
t' \right)})\, , & a>1\, ,
\end{array}\right.
\end{eqnarray}
where $\hyp(\alpha,\beta;\gamma;\cdot)$ denotes the Gauss
hypergeometric function \cite{slater:1960}.

\subsection{Piecewise linear periodic driving}
\label{subsec:IIb} We now introduce the setup for the operational
cycle of the engine under periodic driving. Within a given period,
two branches with linear time-dependence of the state energies are
considered with different velocities. Starting from the value
$h_{1}$, the energy $E_{1}(t)$ linearly increases in the first
branch until it attains the value $h_{2}>h$, at time $t_{+}$ and
in the second branch, the energy $E_{1}(t)$ linearly decreases
towards its original value $h_{1}$ in a time $t_{-}$ (see
figure~1). We always assume $E_{2}(t)=-E_{1}(t)$, i.e.\
\begin{equation}
\label{E1def} E_{1}(t)=-E_2(1)=\left\{
\begin{array}{ll}
h_{1}+\frac{\displaystyle h_{2}-h_{1}}{\displaystyle t_{+}}\,t \,,
& t\in \left[0,t_{+}\right]\, ,\\[1ex]
h_{2}-\frac{\displaystyle h_{2}-h_{1}}{\displaystyle
t_{-}}\,(t-t_{+}) \, , & t\in \left[ t_{+},t_{+}+t_{-} \right] \,
\, .
\end{array} \right.
\end{equation}
This pattern will be periodically repeated, the period being $t_{\rm
  p} = t_{+} + t_{-}$.

As the second ingredient, we need to specify the temperature
schedule. The two-level system will be alternately exposed to a
hot and a cold reservoir, which means that the function $\beta(t)$
in equation~(\ref{transferrates}) will be a piecewise constant
periodic function. During the first branch, it assumes the value
$\beta_{+}$, during the second branch it attains the value
$\beta_{-}$.

This completes the description of the model. Any quantity describing
the engine's performance can only depend on the parameters $h_{1}$,
$h_{2}$, $\beta_{\pm}$, $t_{\pm}$, and $\nu$. In the following we will
focus on the characterization of the limit cycle, which the engine
will approach at long times after a transient period.

We start from the general solution (\ref{PMEsolution}) of the master
equation (\ref{pauliequation}). Owing to the Chapman-Kolmogorov
condition (\ref{Rchapmankolmogorov}), the propagator within the cycle
is
\begin{equation}
\label{LimitCyclePropagatorR} \Rc(t)=\left\{
\begin{array}{ll}
\Rp(t)\, , & t\in \left[ 0,t_{+} \right]\, ,\\[1ex]
\Rm(t)\Rp(t_+)\, , & t\in \left[ t_{+}, \tper
\right]\, .
\end{array} \right.
\end{equation}
Here the matrixes $\mathbb{R}_{\pm}(t)$ evolve the state vector within
the respective branches and have the form
\begin{eqnarray}
\fl \label{Rpmdef} \Rp(t)= \J - \frac{1}{2}
\matr{\phantom{-}1}{-1}{-1}{\phantom{-}1} \left [1 - \exp{\left(-
\nu t \right)} \right ]
+ \frac{1}{2} \matr{-1}{-1}{\phantom{-}1}{\phantom{-}1} \xi_{+}(t)\, , \\
\fl \label{Rmdef} \Rm(t)= \J - \frac{1}{2}
\matr{\phantom{-}1}{-1}{-1}{\phantom{-}1} \left\{1 - \exp{\left[-
\nu (t - t_+) \right]} \right\} + \frac{1}{2}
\matr{-1}{-1}{\phantom{-}1}{\phantom{-}1} \xi_{-}(t)\, ,
\end{eqnarray}
where
\begin{eqnarray}
\fl \label{xiplus}
 \xi_{+}(t) = \nu \int_{0}^{t} \mathrm{d}\tau\, 
\exp{\left[- \nu (t - \tau)\right]} 
\tanh{\left \{ \beta_+ \bigg [ h_1 + \frac{h_2 - h_1}{t_+} \tau \bigg
    ] \right \}}\, , \\
\fl \label{ximinus} \xi_{-}(t) = \nu \int_{t_+}^{t}
\mathrm{d}\tau\, \exp{\left[- \nu (t - \tau)\right]} \tanh{\left
\{ \beta_- \bigg [ h_2 - \frac{h_2 - h_1}{t_-} (\tau - t_+) \bigg
] \right \}}\, .
\end{eqnarray}
Notice, both propagators $\mathbb{R}_{+}(t)$ and
$\mathbb{R}_{-}(t)$ are given by the generic propagator
(\ref{PMEsolution}). In order to get $\mathbb{R}_{+}(t)$, we
replace in equation~(\ref{xidef}) the initial position of the
first energy $h$ by $h_{1}$, the driving velocity $v$ by
$v_{+}=(h_{2}-h_{1})/t_{+}$, and we set $t'=0$. Analogously, the
propagator $\mathbb{R}_{-}(t)$ follows from the generic
propagator, if we replace $h$ by $h_{2}$, $v$ by
$v_{-}=(h_{1}-h_{2})/t_{-}$, and $t'$ by $t_{+}$.
%%%%%%%%%%%%%%%%%%%%%%%%%%%%%%%%%%%%%%%%%%%%%%%%%%%%%%%%%%%%%%%%%%%%%%%%
%%%   Fig 1: Limit cycle                                             %%%
%%%%%%%%%%%%%%%%%%%%%%%%%%%%%%%%%%%%%%%%%%%%%%%%%%%%%%%%%%%%%%%%%%%%%%%%
\begin{figure}[ht]
\begin{center}
\includegraphics[width=1.0\textwidth]{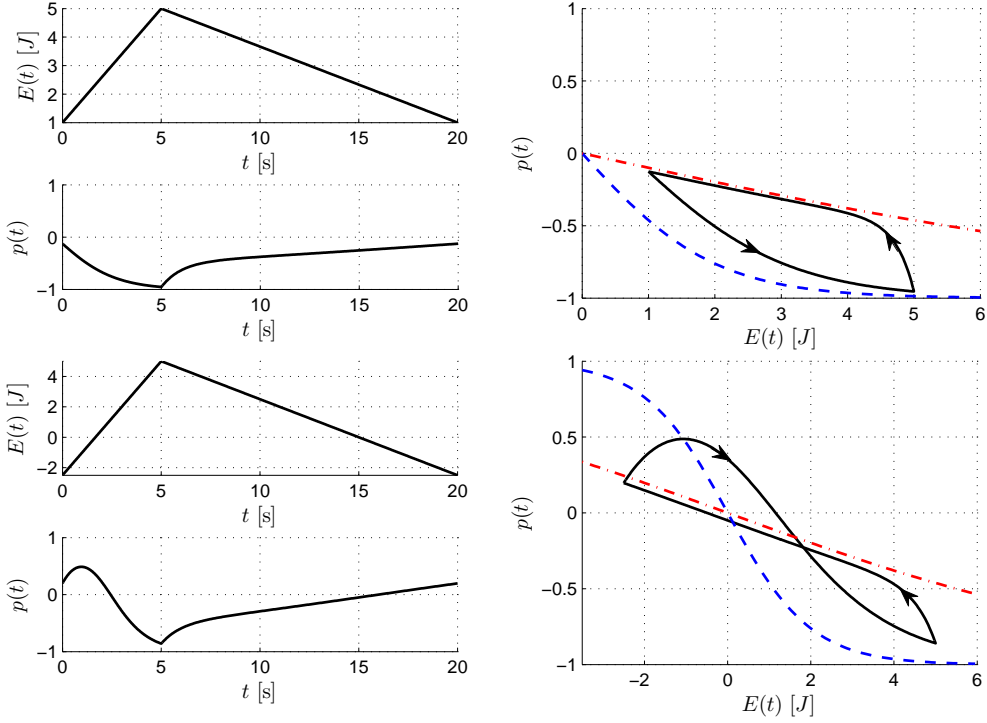}
\end{center}
\caption{The limit cycle for the two-stroke engine. The three
graphs in the upper panel illustrate the case where
$h_{2}>h_{1}>0$ and the energy levels do not cross during their
driving. On the left side we
  show $E(t)=E_{1}(t)=-E_{2}(t)$ and the response
  $p(t)=p_{1}(t)-p_{2}(t)$. On the right hand side the parametric plot
  of the limit cycle in the $p\!-\!\!E$ plane is displayed. The cycle
  starts in the upper vertex and proceeds counterclockwise, c.f.\ the
  arrows. The dashed and the dot-dashed curves show the equilibrium
  isotherms corresponding to the baths during the first and the second
  stroke, respectively. The parameters are: $h_{1}=1\;\mathrm{J}$,
  $h_{2}=5\;\mathrm{J}$, $t_{+}=5\;\mathrm{s}$,
  $t_{-}=15\;\mathrm{s}$, $\beta_{+}=0.5\;\mathrm{J}^{-1}$,
  $\beta_{-}=0.1\;\mathrm{J}^{-1}$, $\nu = 1\; \mathrm{s}^{-1}$. The
  three graphs in the lower panel depict the case where
  $h_{1}<0<h_{2}$ and the energies cross twice during the cycle.
  Except $h_{1}=-2.5\;\mathrm{J}$, all parameters are as above.}
\label{fig1}
\end{figure}
%%%%%%%%%%%%%%%%%%%%%%%%%%%%%%%%%%%%%%%%%%%%%%%%%%%%%%%%%%%%%%%%%%%%%%%%
%%%           End of Fig 1                                           %%%
%%%%%%%%%%%%%%%%%%%%%%%%%%%%%%%%%%%%%%%%%%%%%%%%%%%%%%%%%%%%%%%%%%%%%%%%

The system state probabilities at the ends of the periods form a
Markov chain and we are interested in its fixed point behavior. If we
take the stationary state as the initial condition, the system
revisits this special state at the end of the limit cycle. Therefore
it suffices to solve the eigenvalue problem
$\mathbb{R}_{-}(t_{-})\mathbb{R}_{+}(t_{+})\,\ket{p^{{\rm
      stat}}}=\ket{p^{{\rm stat}}}$. Solving the algebraic equation,
the fixed point probabilities $p_{i}^{\rm stat}$ at the beginning (or
end) of the limit cycle are
\begin{eqnarray}
p_{1}^{\rm stat}=1-p_{2}^{\rm stat}=\frac{1}{2}\left[1-
\frac{\xi_{+}(t_{+})\exp\left(-\nu
t_{-}\right)+\xi_{-}(t_{-})}{1-\exp\left(-\nu t_{{\rm
p}}\right)}\right]. 
\label{p12stat}
\end{eqnarray}
These probabilities, and hence also the specific form of the limit
cycle, depend solely on the model parameters.

We now put aside the transitory regime and we focus entirely on
the limit cycle. Generally speaking, the parametric plot of the
occupation difference $p(t)\equiv p_{1}(t)-p_{2}(t)$ (the
response) versus the energy of the first level $E_{1}(t)$ (the
driving) exhibits two possible forms which are exemplified in
figure~1. First, we have a one-loop form which is oriented either
clockwise or anticlockwise. For clockwise orientation, the work
done by the engine on the environment during the limit cycle is
negative, while for counter-clockwise orientation it is positive.
Secondly, we can obtain a two-loops shape exhibiting again either
positive or negative work on the environment. Slowing down the
driving, the branches gradually approach the corresponding
equilibrium isotherms $p_{\pm}(E)=-\tanh(\beta_{\pm}E/2)$. We
postpone the further discussion to the section~\ref{sec:IV}.

\section{Probability densities for work and heat}
\label{sec:III} Heuristically, the underlying time-inhomogeneous
Markov process $\mathsf{D}(t)$ can be conceived as an ensemble of
individual realizations (sample paths). A realization is specified
by a succession of transitions between the two states. If we know
the number $n$ of the transitions during a path and the times
$\{t_{k}\}_{k=1}^{n}$ at which they occur, we can calculate the
probability that this specific path will be generated. A given
paths yields a unique value of the microscopic work done on the
system. For example, if the system is known to remain during the
time interval $[t_{k},t_{k+1}]$, $t_{k+1}\ge t_{k}$, in the $i$th
state, the work done on the system during this time interval is
simply $E_{i}(t_{k+1})-E_{i}(t_{k})$. Accordingly the probability
of the paths gives the probability of the work. Viewed in this
way, the work itself is a stochastic process and we denote it as
$\mathsf{W}(t)$. We are interested in its probability density
$\rho(w,t)=\langle\,\delta(\mathsf{W}(t)-w)\,\rangle$, where
$\langle\,\ldots\,\rangle$ denotes an average over all possible
paths.

As a technical tool we introduce the \emph{augmented process}
$\left\{\mathsf{W}(t),\,\mathsf{D}(t)\right\}$ which
simultaneously reflects both the work variable and the state
variable of the Markov process. The augmented process is again a
time non-homogeneous Markov process. Actually, if we know at a
fixed time $t'$ both the present state variable $j$ and the work
variable $w'$, then the subsequent probabilistic evolution of the
state and work is completely determined. The work done during a
time period $[t',t]$, where $t > t'$, simply adds to the present
work $w'$ and it only depends on the succession of the states
after the time $t'$. And this succession by itself, as we know
from section~\ref{sec:II}, cannot depend on the dynamics before
time $t'$.

The one-time properties of the augmented process will be described by
the functions
\begin{eqnarray}
\fl G_{ij}(w,t\,|\,w',t')=\lim_{\epsilon\rightarrow 0}\frac{ {\rm
Prob}\left\{\,\mathsf{W}(t)\in(w,w+\epsilon)\,{\rm and}\,\mathsf
{D}(t)=i\,|\,\mathsf{W}(t')=w'\,{\rm
and}\,\mathsf{D}(t')=j\,\right\}} {\epsilon}\, , \nonumber \\
\label{gdef1}
\end{eqnarray}
where $i,j=1,2$. We represent them as the matrix elements of a single
two-by-two matrix ${\mathbb G}(w,w';t,t')$,
\begin{equation}
\label{gdef2} G_{ij}(w,t\,|\,w',t')=\langle\,i\,|\,
{\mathbb G}(w,t\,|\,w',t')\,|\,j\,\rangle\,\,.
\end{equation}
Using this matrix notation, the Chapman-Kolmogorov condition for the
augmented process assumes the form
\begin{equation}
\label{Gchapmankolmogorov} {\mathbb G}(w,t\,|\,w',t')=
\int\,dw''\,{\mathbb G}(w,t\,|\,w'',t'')\,{\mathbb G}(w'',t''\,|\,w',t')\,\,.
\end{equation}
Here the matrix multiplication on the right hand side amounts for the
summation over the intermediate states at the time $t''$, and the
integration runs over all possible intermediate values of the work
variable $w''$. The equation is valid for any intermediate time
$t''\in[t',t]$. Similarly to the preceding section~\ref{sec:II}, the
Chapman-Kolmogorov condition can be used to connect two different
propagators describing the time evolution of the augmented process
within two branches of the driving cycle. Before we address this
point, we focus on the generic situation.

We need an equation which controls the time dependence of the
propagator ${\mathbb G}(w,t\,|\,w',t')$ and which plays the same role
as the rate equation (\ref{pauliequation}) in the case of the simple
two-state process. It reads
\begin{eqnarray}
\fl \frac{\partial}{\partial t}{\mathbb
G}(w,t\,|\,w',t')=-\left\{\, \frac{\partial}{\partial w}
\left(\begin{array}{cc}
\frac{d\,E_{1}(t)}{dt}&0\\
0&\frac{d\,E_{2}(t)}{dt}
\end{array}\right)
+\left(\begin{array}{cc}
\phantom{-}\lambda_{1}(t)&-\lambda_{2}(t)\\
-\lambda_{1}(t)&\phantom{-}\lambda_{2}(t)
\end{array}\right)\,\right\}\,{\mathbb
G}(w,t\,|\,w',t'), \nonumber \\
\label{Gdynequation}
\end{eqnarray}
where the initial condition is ${\mathbb
  G}(w,t'\,|\,w',t')=\delta(w-w'){\mathbb I}$. This is a hyperbolic
system of four coupled partial differential equations with
time-dependent coefficients. It can be derived in several ways. For
example, as explained in reference \cite{Imparato/Peliti:2005c}, one
considers at the time $t$ the family of all realizations, which
display at that time the work in the infinitesimal interval $(w,w+dw)$
and, simultaneously, which occupy a given state. During the
infinitesimal time interval $(t,t+dt)$, the number of such paths can
change due to two reasons. First, while residing in the given state,
some paths enter (leave) the set, because the energy levels move and
an additional work has been done. Secondly, some paths can enter
(leave) the described family because they jump out of (into) the
specified state. These two contributions correspond to the two terms
on the right hand side of equation~(\ref{Gdynequation}). Another
derivation \cite{Chvosta/etal:2007} is based on an explicit
probabilistic construction of all possible paths and their respective
probabilities.

Similar reasoning holds for the random variable $\RW{Q}(t)$ describing
the heat accepted by the system from the environment, and for the
internal energy $\RW{U}(t)$. The variable $\RW{Q}(t)$ is described by
the propagator $\matice{K}(q,t\,|\,q',t')$ with the matrix elements
\begin{eqnarray}
\label{KijMeaning} \fl
K_{ij}(q,t\,|\,q',t')=\lim_{\epsilon\rightarrow 0}\frac
{\mathop{\mathrm{Prob}}\big
\{\mathsf{Q}(t)\in(q,q+\epsilon)\,\wedge\,\mathsf{D}(t)=i\,|\,\mathsf{Q}(t')=q'\,\wedge\,\mathsf{D}(t')=j\big\}}
{\epsilon}\,.
\end{eqnarray}
It turns out that there exists a simple connection between the heat
propagator and the work propagator ${\mathbb G}(w,t\,|\,w',t')$. Since
for each path, heat $q$ and work $w$ are connected by the first law of
thermodynamics, we have $q=E_i(t)-E_j(t')-w$ for any path which has
started at time $t'$ in the state $i$ and which has been found at time
$t$ in the state $j$. Accordingly,
\begin{eqnarray}
\fl \label{Qij} \matice{K}(q,t\,|\,q',t')=\matr
{g_{11}(u_{11}(t,t')-q,t\,|\,q',t')}
{g_{12}(u_{12}(t,t')-q,t\,|\,q',t')}
{g_{21}(u_{21}(t,t')-q,t\,|\,q',t')}
{g_{22}(u_{22}(t,t')-q,t\,|\,q',t')} \,\,,
\end{eqnarray}
where $u_{ij}(t,t')=E_i(t)-E_j(t')$. This relation can be written in
the form of the symmetry relation
\begin{equation}
\label{symmetryrelation}
G_{ij}(u_{ij}(t,t')/2+q,t\,|\,q',t')=K_{ij}(u_{ij}(t,t')/2-q,t\,|\,q',t')\,\,.
\end{equation}

\subsection{Generic case--linear driving protocol}
\label{subsec:IIIa} For the linear driving protocol
$E_1(t)=h+v(t-t')=-E_2(t)$ the first term in the curly brackets in
equation~(\ref{Gdynequation}) is time-independent. As for the second
term, we use again the Glauber rates (\ref{transferrateslinear}).
Thereby the evolution equation (\ref{Gdynequation}) assumes the form

\begin{eqnarray}
\fl \label{WPDEquationGeneric} \Pder{}{t} {\mathbb
G}(w,t'\,|\,w',t')= -\left \{v \Pder{}{w} \JmJ \right. \\
\left. +
\frac{\nu}{1+c\exp{[-\Omega(t-t')}]}
\matr{1}{-c\exp{[-\Omega(t-t')]}}{-1}{c\exp{[-\Omega(t-t')]}}
\right\} {\mathbb G}(w,t'\,|\,w',t')\, , \nonumber
\end{eqnarray}
with the parameters $c$ and $\Omega$ introduced in connection with
equation~(\ref{transferrateslinear}).

We shall now employ the method described in
reference~\cite{Chvosta/etal:2007} by taking the double Laplace
transformation with respect to the variables $t$ and $w$. As shown in
reference~\cite{Chvosta/etal:2007}, a special difference equation
results, which can be solved exactly. Moreover, it is possible to
carry out the final double inverse Laplace transformation. However, in
\cite{Chvosta/etal:2007}, only the case $E_1(0)=E_2(0)=0$ has been
studied. In the present context we need the solution for a general
initial energy difference $2h$. It turns out that such generalization
represents a nontrivial task. The constant $c$ cannot be simply scaled
off because it enters only the second term on the right hand side of
equation~(\ref{Gdynequation}). In order to overcome this difficulty,
we had to modify the procedure from
reference~\cite{Chvosta/etal:2007}. However, in view of the specific
topic of the present paper, we refrain from giving the technical
details and we proceed with the description of the final result.

For the presentation of the result it is convenient to introduce the
reduced work variable $\eta=\eta(w,w')=2\beta(w-w')$ and the reduced
time variable $\tau=\tau(t,t')=\Omega(t-t')$. Moreover, it is helpful
to use the abbreviations
\begin{equation}
\label{abbr}
x=\exp\left[-\frac{\tau+\eta}{2}\right]
\,\,,\,\,
y=\exp\left[-\frac{\tau - \eta}{2}\right]
\,\,,\,\,
\phi=-c\,\frac{1-x}{1+c x}\,\frac{1-y}{1+c y}\,\,.
\end{equation}
For $v>0$, the result is
\begin{eqnarray}
%%%%%%%%%%%
%%% g11 %%%
%%%%%%%%%%%
\fl \frac{1}{2\beta}G_{11}(\eta,\tau\,|\,\eta',\tau')= \left[
\frac{(1+c)\exp(-\tau)}{1+c\exp(-\tau)}\right]^{a}\,(\tau-\eta)+
\Theta(\tau+\eta)\Theta(\tau-\eta)\,\frac{ac}{2}x^{a}(1-x)y \nonumber\\
\label{g11}
\hspace*{-2cm}\times
\left[-\frac{\hyp\!(1+a,-a;1;\phi)}{(1+cx)^{1+a}(1+cy)^{1-a}}+
(1+a)(1+c)(1+cxy)
\frac{\hyp\!(2+a,1-a;2;\phi)}{(1+cx)^{2+a}(1+cy)^{2-a}}\right],\\[1ex]
%%%%%%%%%%%
%%% g12 %%%
%%%%%%%%%%%
\fl \label{g12} \frac{1}{2\beta}G_{12}(\eta,\tau\,|\,\eta',\tau')=
\frac{1}{2}\Theta(\tau+\eta)\Theta(\tau-\eta)\,
acx^{a}y\,\frac{\hyp\!(a,1-a;1;\phi)}{(1+cx)^{a}(1+cy)^{1-a}}\,,\\[1ex]
%%%%%%%%%%%
%%% g21 %%%
%%%%%%%%%%%
\fl \label{g21} \frac{1}{2\beta}G_{21}(\eta,\tau\,|\,\eta',\tau')=
\frac{1}{2}\Theta(\tau+\eta)\Theta(\tau-\eta)\,
ax^{a}\,\frac{\hyp\!(1+a,-a;1;\phi)}{(1+cx)^{1+a}(1+cy)^{-a}}\,,\\[1ex]
%%%%%%%%%%%
%%% g22 %%%
%%%%%%%%%%%
\fl \frac{1}{2\beta}G_{22}(\eta,\tau\,|\,\eta',\tau')=
\left[\frac{1+c\exp(-\tau)}{1+c}\right]^{a}\,(\tau+\eta)+
\Theta(\tau+\eta)\Theta(\tau-\eta)\,\frac{ac}{2}x^{a}(1-y)\nonumber\\
\label{g22}
\hspace*{-2cm}\times
\left[+\frac{\hyp\!(a,1-a;1;\phi)}{(1+cx)^{1+a}(1+cy)^{1-a}}-
(1-a)(1+c)(1+cxy)
\frac{\hyp\!(1+a,2-a;2;\phi)}{(1+cx)^{2+a}(1+cy)^{2-a}}\right].
\end{eqnarray}
Here $\delta(\cdot)$ is the Dirac delta-function, and $\Theta(\cdot)$
is the Heaviside unit step function. The solution for $v<0$ follows
from interchanging the indices $1$ and $2$ in
equations~(\ref{g11})-(\ref{g22}). If $h=0$, then $c=1$, and our
results coincide with the formulae (49)--(52) in
reference~\cite{Chvosta/etal:2007}.

\subsection{Piecewise linear periodic driving}
\label{subsec:IIIb} The generic result (\ref{g11})-(\ref{g22})
immediately yields the work and heat propagators for the individual
branches in the protocol according to equation~(\ref{E1def}). We
simply carry out the replacements described in the text following
equation~(\ref{xiplus}). We denote the corresponding matrices as
$\matice{G}_{\pm}(w,w',t)$ and $\matice{K}_{\pm}(w,w',t)$. Then the
Chapman-Kolmogorov condition (\ref{Gchapmankolmogorov}) yields the
propagator
\begin{eqnarray}
\fl \label{LimitCyclePropagatorG}
\matice{G}_{\mathrm{p}}(w,t)=\left\{ \begin{array}{ll}
\matice{G}_+(w,0,t)\,, & t\in[0,t_{+}]\,,\\[1ex]
\displaystyle
\int_{-(h_2-h_1)}^{h_2-h_1}\der{w'}\,
\matice{G}_-(w,w',t)\matice{G}_+(w',0,t_+)\,,
& t\in[t_{+},\tper]\,.
\end{array} \right.
\end{eqnarray}
As demonstrated above, the heat propagator
$\matice{K}_{\mathrm{p}}(w,t)$ for the limit cycle is connected with
the work propagator $\matice{G}_{\mathrm{p}}(w,t)$ through simple
shifts of the independent variable $w$. Specifically, we get
$\langle\,i\,|\,\mathbb{K}_{\rm
  p}(q,t)\,|\,j\,\rangle=\langle\,i\,|\,\mathbb{G}_{\rm
  p}(u_{ij}(t)-q,t)\,|\,j\,\rangle$ with $u_{21}(t)=-u_{12}(t)$,
$u_{22}(t) = - u_{11}(t)$, and
\begin{eqnarray}
\fl \label{ufirststroke} u_{11}(t)&=&\frac{h_2 -
h_1}{t_+}t,\qquad\qquad\qquad\;\;
u_{12}(t)=2h_1+\frac{h_2-h_1}{t_+}t,
\qquad\qquad\quad \;\, t\in[0,t_+],\\
\fl \label{usecondstroke} u_{11}(t)&=&(h_2 - h_1)\left(1-\frac{t -
t_+}{t_-}\right),\; u_{12}(t)=h_2+h_1-\frac{h_2-h_1}{t_-}(t-t_+),
\quad t\in[t_+,t_{\rm p}].
\end{eqnarray}

In the last step we take into account the initial condition
(\ref{p12stat}) at the beginning of the limit cycle and we sum over
the final states of the process $\mathsf{D}(t)$. Then the probability
density for the work done on the system during the limit cycle reads
\begin{equation}
\label{rhocycle}
\rho_{\rm p}(w,t)=\sum_{i=1}^{2}\langle\,i\,|\,{\mathbb G}_{\rm p}(w,t)|\,p^{\rm stat}\,\rangle\,\,.
\end{equation}
Similarly, the probability density for the head accepted within the
limit cycle is
\begin{equation}
\label{rhoHeatCycle}
\chi_{\mathrm{p}}(q,t)=\sum_{i=1}^2\langle\,i\,|\,{\mathbb K}_{\rm p}(q,t)|\,p^{\rm stat}\,\rangle\,\,.
\end{equation}
These two functions represent the main results of the present Section.
They are illustrated in figures~2-4. We discuss their main features in
section~\ref{sec:V}.
%%%%%%%%%%%%%%%%%%%%%%%%%%%%%%%%%%%%%%%%%%%%%%%%%%%%%%%%%%%%%%%%%%%%%%%%
%%%   Fig 2: Time resolved work density                              %%%
%%%%%%%%%%%%%%%%%%%%%%%%%%%%%%%%%%%%%%%%%%%%%%%%%%%%%%%%%%%%%%%%%%%%%%%%
\begin{figure}[!ht]
\centering
\includegraphics[width=1.0\textwidth]{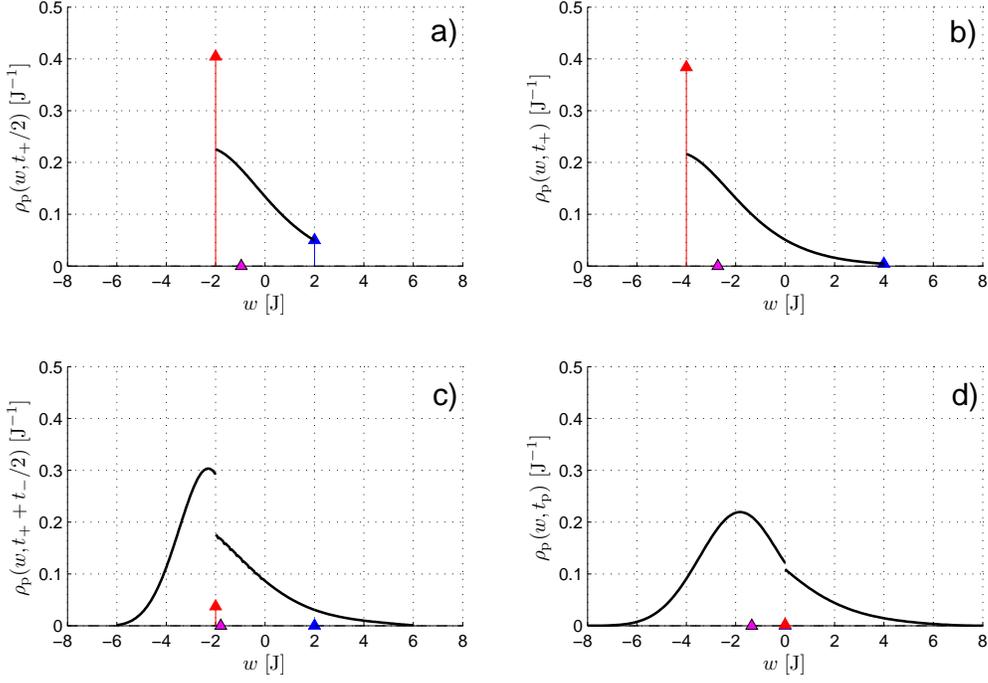}
\caption{The probability density $\rho_{\mathrm{p}}(w,t)$ as a
  function of the work $w$ for the same parameters as in figure~1 (with
  positive $h_{1}$): a) $t=\frac{1}{2}t_{+}$ (middle of the first
  stroke), b) $t=t_{+}$ (end of the first stroke), c)
  $t=t_{+}+\frac{1}{2}t_{-}$ (middle of the second stroke), and d)
  $t=t_{+}+t_{-}$ (end of the limit cycle). The triangle on the work
  axis marks the mean work $W(t)$ at the corresponding times. The
  singular parts of $\rho_{\mathrm{p}}(w,t)$ are marked by arrows,
  where the arrow heights equal the weights of the corresponding delta
  functions [for example, in panel a), the left arrow height gives the
  probability that the system is initially in the second state and
  remains in it between the beginning of the cycle and the time
  $t=\frac{1}{2}t_{+}$; then the work done on the system equals
  $-\frac{1}{2}(h_2-h_1)$].
\label{fig2}
}
\end{figure}
%%%%%%%%%%%%%%%%%%%%%%%%%%%%%%%%%%%%%%%%%%%%%%%%%%%%%%%%%%%%%%%%%%%%%%%%
%%%   End of Fig 2                                                   %%%
%%%%%%%%%%%%%%%%%%%%%%%%%%%%%%%%%%%%%%%%%%%%%%%%%%%%%%%%%%%%%%%%%%%%%%%%
%%%%%%%%%%%%%%%%%%%%%%%%%%%%%%%%%%%%%%%%%%%%%%%%%%%%%%%%%%%%%%%%%%%%%%%%
%%%   Fig 3: Time resolved heat density                              %%%
%%%%%%%%%%%%%%%%%%%%%%%%%%%%%%%%%%%%%%%%%%%%%%%%%%%%%%%%%%%%%%%%%%%%%%%%
\begin{figure}[ht]
\begin{center}
\includegraphics[angle=0,width=1.0\textwidth]{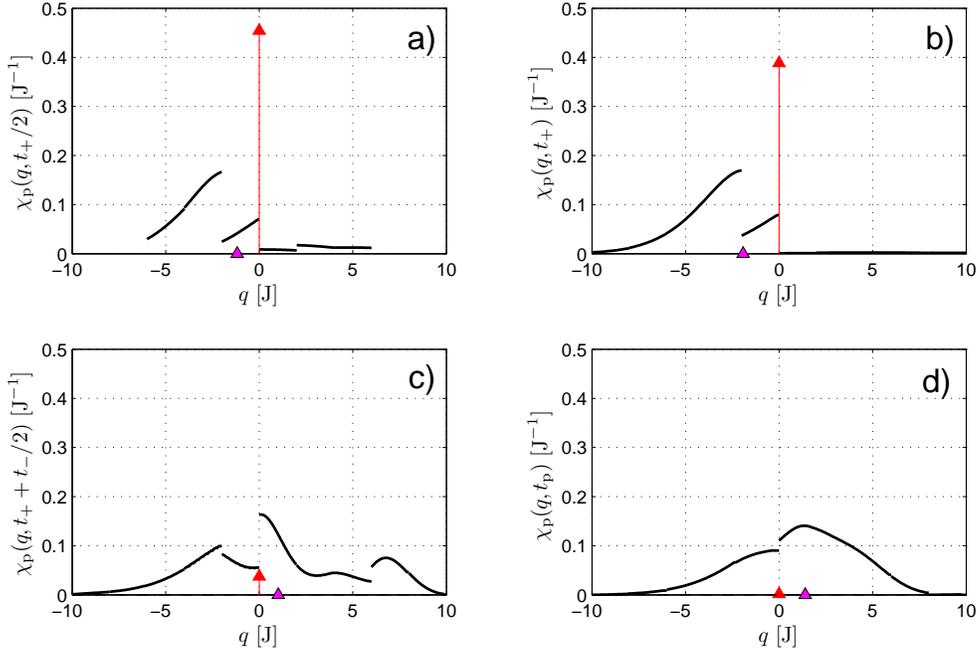}
\caption{The probability density $\chi_{\mathrm{p}}(q,t)$ as a
  function of the heat $q$ and for the same parameters as in figure~1 (with
  positive $h_{1}$): a) $t=\frac{1}{2}t_{+}$ (middle of the first
  stroke), b) $t=t_{+}$ (end of the first stroke), c)
  $t=t_{+}+\frac{1}{2}t_{-}$ (middle of the second stroke), and d)
  $t=t_{+}+t_{-}$ (end of the limit cycle).
  The triangles on the heat axis mark the mean heat $Q(t)$ at the
  corresponding times. The singular parts of $\chi_{\mathrm{p}}(q,t)$
  are marked by the arrow, where the arrow height equals
  the weight of the corresponding delta function. For example, in
  a), the height of the arrow gives the probability that
  there was no transition between the states from the beginning of the
  cycle till the observation time $t=\frac{1}{2}t_{+}$. The heat
  exchanged in this case is zero.
\label{Time_Heat}
}
\end{center}
\end{figure}
%%%%%%%%%%%%%%%%%%%%%%%%%%%%%%%%%%%%%%%%%%%%%%%%%%%%%%%%%%%%%%%%%%%%%%%%
%%%   End of Fig 3                                                   %%%
%%%%%%%%%%%%%%%%%%%%%%%%%%%%%%%%%%%%%%%%%%%%%%%%%%%%%%%%%%%%%%%%%%%%%%%%
%%%%%%%%%%%%%%%%%%%%%%%%%%%%%%%%%%%%%%%%%%%%%%%%%%%%%%%%%%%%%%%%%%%%%%%%
%%%   Fig 4: Work and heat densities for the limit cycle             %%%
%%%%%%%%%%%%%%%%%%%%%%%%%%%%%%%%%%%%%%%%%%%%%%%%%%%%%%%%%%%%%%%%%%%%%%%%
\begin{figure}[!ht]
\centering
\includegraphics[width=1.0\textwidth]{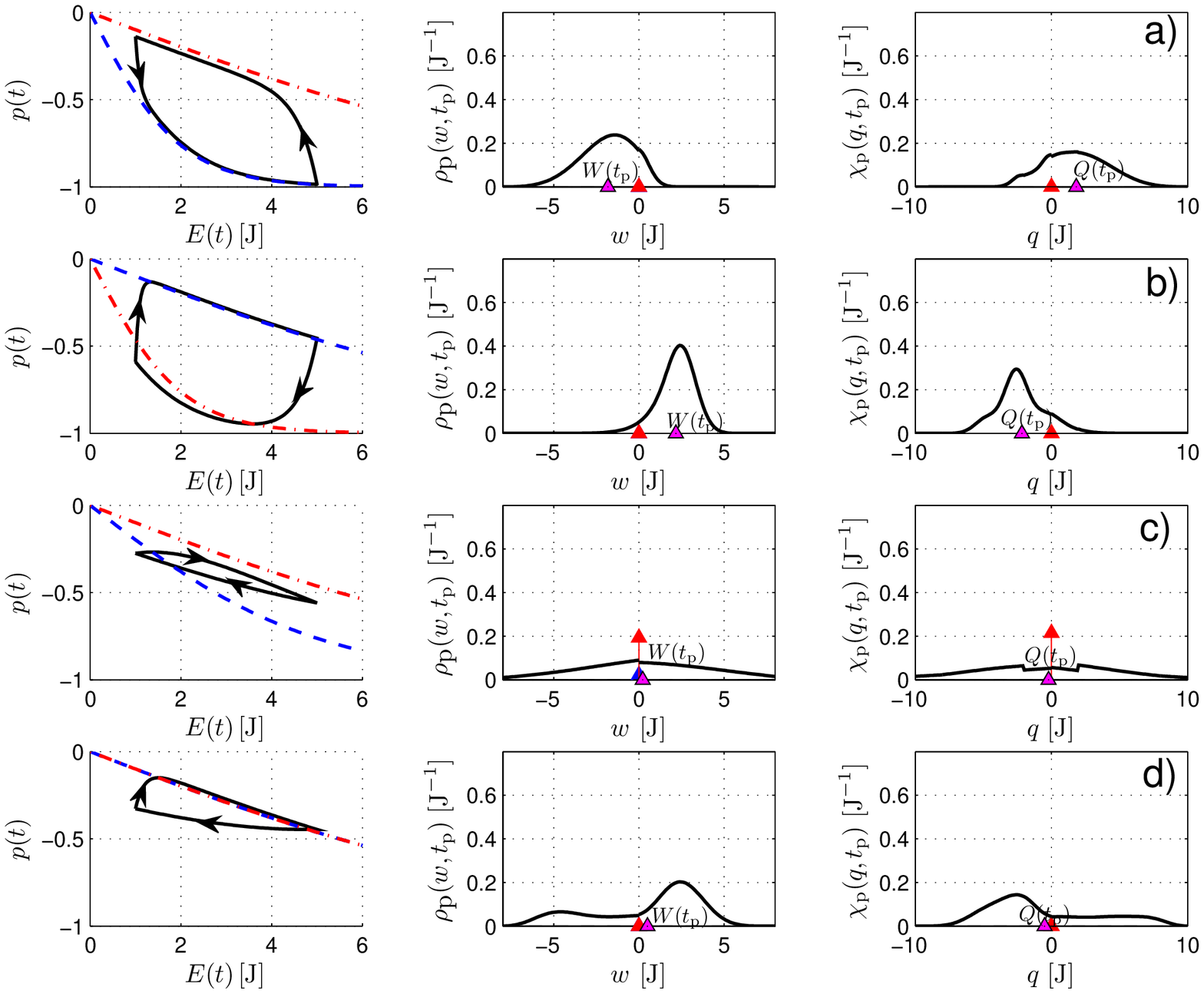}
\caption{ Probability densities $\rho_{\rm p}(w,\tper)$ and $\chi_{\rm
    p}(q,\tper)$ for the work and heat for four representative sets of
  the engine parameters. For every set we show also the limit cycle in
  the $p\!-\!\!E$ plane, where the corresponding equilibrium isotherms
  are marked by dashed (first stroke) and dot-dashed (second stroke)
  lines. In all cases we choose $h_{1}=1\;\mathrm{J}$,
  $h_{2}=5\;\mathrm{J}$, and $\nu=1\;\mathrm{s}^{-1}$. The remaining
  parameters are a) $t_{+}=50\;\mathrm{s}$, $t_{-}=10\;\mathrm{s}$,
  $\beta_{+}=0.5\;\mathrm{J}^{-1}$, $\beta_{-}=0.1\;\mathrm{J}^{-1}$
  (bath of the first stroke is colder than of the second stroke),
  b) $t_{+}=50\;\mathrm{s}$, $t_{-}=10\;\mathrm{s}$,
  $\beta_{+}=0.1\;\mathrm{J}^{-1}$, $\beta_{-}=0.5\;\mathrm{J}^{-1}$
  (exchange of $\beta_{+}$ and $\beta_{-}$ as compared to case a), leading
  to a change of the traversing of the cycle from counter-clockwise to
  clockwise and a sign reversal of the mean values
  $W(\tper)\equiv\langle\,\mathsf{W}(\tper)\,\rangle$ and
  $Q(\tper)\equiv\langle\,\mathsf{Q}(\tper)\,\rangle$), c)
  $t_{+}=2\;\mathrm{s}$, $t_{-}=2\;\mathrm{s}$,
  $\beta_{+}=0.2\;\mathrm{J}^{-1}$, $\beta_{-}=0.1\;\mathrm{J}^{-1}$
  (a strongly irreversible cycle traversed clockwise with positive
  work), d) $t_{+}=20\;\mathrm{s}$, $t_{-}=1\;\mathrm{s}$,
  $\beta_{\pm}=0.1\;\mathrm{J}^{-1}$ (no change in temperatures, but
  large difference in duration of the two strokes; $W(\tper)$ is
  necessarily positive).
\label{fig3}
}
\end{figure}
%%%%%%%%%%%%%%%%%%%%%%%%%%%%%%%%%%%%%%%%%%%%%%%%%%%%%%%%%%%%%%%%%%%%%%%%
%%%   End of Fig 4                                                   %%%
%%%%%%%%%%%%%%%%%%%%%%%%%%%%%%%%%%%%%%%%%%%%%%%%%%%%%%%%%%%%%%%%%%%%%%%%

\section{Engine performance}
\label{sec:IV} As shown in section~\ref{sec:II}, the occupation
probabilities during the limit cycle are $\mathbb{R}_{{\rm
    p}}(t)\,\ket{p^{{\rm stat}}}$ with $\mathbb{R}_{{\rm p}}(t)$ given
by equation~(\ref{LimitCyclePropagatorR}). These probabilities are
ensemble averaged quantities and cannot describe fluctuations of the
engine's performance. But they render the energetics in terms of mean
values as we discuss now.

During the limit cycle, the internal energy $U(t)=\sum_{i=1}^{2}
E_i(t) p_i(t)$ changes as
\begin{eqnarray}
\fl \label{eq:dU} \frac{\mathrm{d}}{\mathrm{d}
t}U(t)=\sum_{i=1}^{2}E_{i}(t)\frac{\mathrm{d}}{\mathrm{d}
t}p_{i}(t)+\sum_{i=1}^{2}p_{i}(t)\frac{\mathrm{d}}{\mathrm{d}
t}E_{i}(t)= \frac{\mathrm{d}}{\mathrm{d} t}\left[Q(t)+W(t)\right],
\quad \,\,\,t\in[0,t_{{\rm p}}]\,.
\end{eqnarray}
Here $Q(t)\equiv\langle\,\mathsf{Q}(t)\,\rangle$ is the mean heat
received from the reservoirs during the period between the beginning
of the limit cycle and the time $t$. Analogously
$W(t)\equiv\langle\,\mathsf{W}(t)\,\rangle$ is the mean work done on
the system from the beginning of the limit cycle till the time $t$. If
$W(t)<0$, the positive work $-W(t)$ is done by the system on the
environment. Therefore the \emph{oriented} areas enclosed by the limit
cycle in figure~1 and in figure~4 represent the work
$\Wout\equiv-W(\tper)$ done by the engine on the environment per
cycle. These areas approach maximal absolute values in the
quasi-static limit. The internal energy, being a state function,
fulfills $U(t_{{\rm p}})=U(0)$. Therefore, if the work $\Wout$ is
positive, the same total amount of heat has been transferred from the
two reservoirs during the cycle. The case $\Wout>0$ cannot occur if
both reservoirs would have the same temperature. That the
\emph{perpetuum mobile} is actually forbidden can be traced back to
the detailed balance condition in (\ref{pauliequation}).
%%%%%%%%%%%%%%%%%%%%%%%%%%%%%%%%%%%%%%%%%%%%%%%%%%%%%%%%%%%%%%%%%%%%%%%%
%%%   Fig 5: Thermodynamics: Mean work, mean heat, etc.              %%%
%%%%%%%%%%%%%%%%%%%%%%%%%%%%%%%%%%%%%%%%%%%%%%%%%%%%%%%%%%%%%%%%%%%%%%%%
\begin{figure}[ht]
\begin{center}
\includegraphics[width=0.85\textwidth]{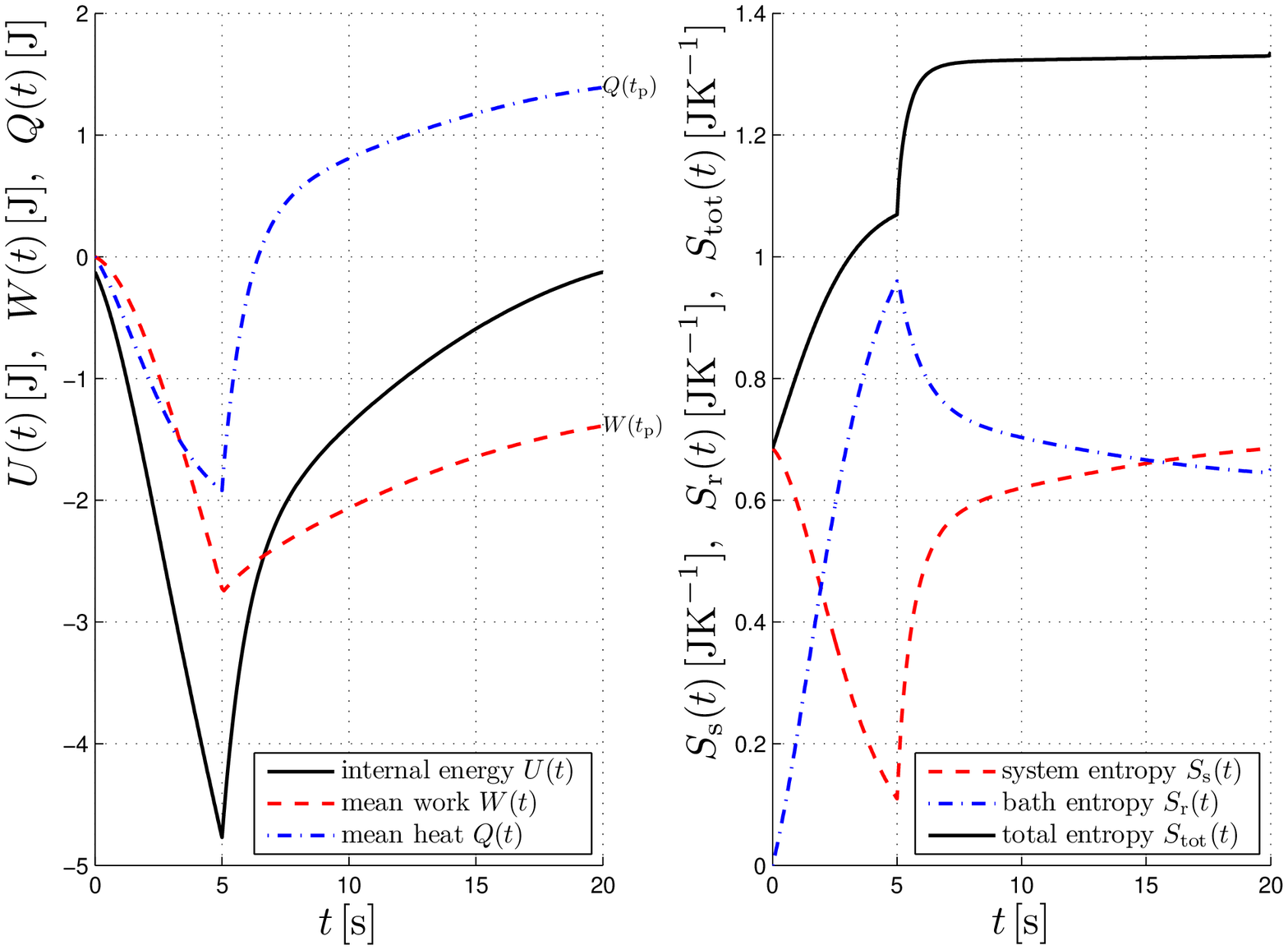}
\end{center}
\caption{Thermodynamic quantities as functions of time during the
  limit cycle for the same set of parameters as in the upper panel of
  figure~1 (positive $h_{1}$). Left panel: internal energy, mean work
  done on the system, and mean heat received from both reservoirs; the
  final position of the mean work curve marks the work done on the
  system per cycle $W(t_{\rm p})$. Since $W(t_{\rm p})<0$, the work
  $W_{\rm out}=-W(t_{\rm p})$ has been done on the environment. The
  internal energy returns to its original value and, after completion
  of the cycle, the absorbed heat $Q(t_{\rm p})$ equals the negative
  work $-W(t_{\rm p})$. Right panel: entropy $S_{\mathrm{s}}(t)$ of
  the system and $S_{\mathrm{r}}(t)$ of the bath, and their sum
  $S_{\rm tot}(t)$; after completing the cycle, the system entropy
  re-assumes its initial value. The difference $S_{\rm tot}(t_{\rm
    p})-S_{\rm tot}(0)>0$ equals the entropy production per cycle. It
  is always positive and quantifies the degree of irreversibility of
  the cycle. \label{fig4} }
\end{figure}
%%%%%%%%%%%%%%%%%%%%%%%%%%%%%%%%%%%%%%%%%%%%%%%%%%%%%%%%%%%%%%%%%%%%%%%%
%%%   End of Fig 5                                                   %%%
%%%%%%%%%%%%%%%%%%%%%%%%%%%%%%%%%%%%%%%%%%%%%%%%%%%%%%%%%%%%%%%%%%%%%%%%

We denote the system entropy at time $t$ as $S_{\mathrm{s}}(t)$, and
the reservoir entropy at time $t$ as $S_{\mathrm{r}}(t)$. They are
given by
\begin{eqnarray}
\label{eq:Ss}
\fl \frac{S_{\mathrm{s}}(t)}{k_\mathrm{B}}=
-\left[p_1(t)\ln p_1(t)+ _2(t)\ln p_2(t)\right],\\
\label{eq:Sr} \fl
\frac{S_{\mathrm{r}}(t)}{k_\mathrm{B}}=
-\beta_{+}\int_{0}^{t_{+}}\der{t'}E_1(t')\Nder{}{t'}[p_1(t')
- p_2(t')]- \beta_{-}\int_{t_{+}}^{t_{{\rm
p}}}\der{t'}E_1(t')\Nder{}{t'}[p_1(t') - p_2(t')].
\end{eqnarray}
Upon completing the cycle, the system entropy re-assumes its value at
the beginning of the cycle. On the other hand, the reservoir entropy
is controlled by the heat exchange. Owing to the inherent
irreversibility of the cycle we observe always a positive entropy
production per cycle, $S_{\mathrm{r}}(t_{{\rm
    p}})-S_{\mathrm{r}}(0)>0$. The total entropy
$S_{\mathrm{tot}}(t)=S_{\mathrm{s}}(t)+S_{\mathrm{r}}(t)$ increases
for any $t\in[0,t_{{\rm p}}]$. The rate of the increase is the larger
the stronger is the representative point in the $p\!-\!\!E$ diagram
deviates from the corresponding equilibrium isotherm (a strong
deviation, e.g., can be seen in the $p\!-\!\!E$ diagram in figure~4c).
Due to the instantaneous exchange of the baths at times $t_{+}$ and
$t_{+}+t_{-}$ in the model considered here, a strong increase of
$S_{\mathrm{tot}}(t)$ always occurs after these time instants. A
representative example of the overall behavior of the thermodynamic
quantities (mean work and heat, and entropies) during the limit cycle
is shown in figure~5.

An important characteristics of the engine is its power output $\Pout$
and its efficiency $\mu$. They are defined as
\begin{equation}
\label{PowerOutputEfficiency}
\Pout\equiv\frac{\Wout}{\tper}\,\,\,,\,\,\,\mu\equiv\frac{\Wout}{\Qin} \,\,,
\end{equation}
where $\Qin$ is the total heat absorbed by the system per cycle. The
performance of the engine characterized by the output work,
efficiency, output power, and entropies from equations~(\ref{eq:Ss})
and (\ref{eq:Sr}) are shown in figure~6 and figure~7.

In figure~6 the performance is displayed as a function of the cycle
duration $t_{\rm p}$ for $t_{+}=t_{-}=\tper/2$. With increasing
$t_{\rm p}$, the output work and the efficiency increase whereas the
output power and the entropy production first increase up to a maximum
and thereafter they decrease when approaching the quasi-static limit
($t_{{\rm p}}\rightarrow\infty$). Notice that the maximum efficiency
and output power occur at different values of $t_{\rm p}$. In
figure~6a) we show also the standard deviation of the output work,
which was calculated from the work probability density
$\rho_{{\mathrm{p}}}(w,\tper)$. Finally, let us note that the values
$\beta_{+}=0.5$ $\mathrm{J}^{-1}$ and $\beta_{-}=0.1$
$\mathrm{J}^{-1}$ used in figure~6 give the Carnot efficiency
$\mu_{\mathrm{C}}=0.8$. This should be compared with the efficiency of
the engine for a long period $t_{\rm p}$, that is, with the value
$\mu\approx 0.6$. As discussed above, the Carnot efficiency cannot be
reached here even for $t_{{\rm p}}\rightarrow\infty$, due to the
immediate temperature changes at times $t_{+}$ and $t_{+}+t_{-}$.

In figure~7 we have fixed $t_{{\rm p}}$ and plotted the behavior as
function of the time asymmetry (or time splitting) parameter
$\Delta=(t_{+}-t_{-})/t_{{\rm p}}$. As can be seen from the upper
three panels in figure~7, there exist also a maximal efficiency and a
maximal output power with respect to a variation of the time asymmetry
parameter (as long as the engine performs work, i.e., $\Wout>0$).
Again, the optimal parameter $\Delta$, where these maxima occur, is
different for the efficiency and output power. In a reversed
situation, considered in the lower three panels in figure~7, where the
work is performed on the engine ($\Wout<0$), minima of the efficiency
and output power occur.
%%%%%%%%%%%%%%%%%%%%%%%%%%%%%%%%%%%%%%%%%%%%%%%%%%%%%%%%%%%%%%%%%%%%%%%%
%%%   Fig 6: Performance versus duration of the limit cycle          %%%
%%%%%%%%%%%%%%%%%%%%%%%%%%%%%%%%%%%%%%%%%%%%%%%%%%%%%%%%%%%%%%%%%%%%%%%%
\begin{figure}[ht]
\begin{center}
\includegraphics[angle=0,width=0.7\textwidth]{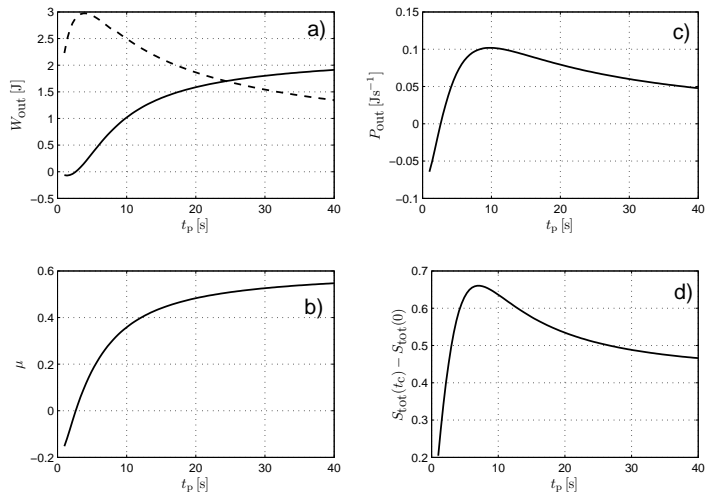}
\caption{The engine performance versus the duration of the limit cycle
  $t_{\pm}$ for $t_{+}=t_{-}={\textstyle\frac{1}{2}}t_{\rm p}$ and
  otherwise the same parameters as in the upper part of figure~1
  (positive $h_{1}$). Both the output work $\Wout$ in a) and the
  efficiency $\mu$ in b) increase with $t_{{\rm p}}$. The output power
  $\Pout$ in c) assumes a maximum at a special cycle duration. The
  dashed line in a) marks the standard deviation of the output work,
  calculated from the work density $\rho_{\rm p}(w,\tper)$. Notice
  that the work fluctuation is comparatively high close to the cycle
  duration, where the maximal output power is found. In the
  long-period limit $t_{{\rm p}}\rightarrow\infty$, the cycle still
  represents a non-equilibrium process (due to the construction of the
  model, see text), and hence the entropy production $S_{\rm
    tot}(t_{\rm p})-S_{\rm tot}(0)$ in d) remains positive,
  approaching a specific asymptotic value.
\label{Performance_period}
}
\end{center}
\end{figure}
%%%%%%%%%%%%%%%%%%%%%%%%%%%%%%%%%%%%%%%%%%%%%%%%%%%%%%%%%%%%%%%%%%%%%%%%
%%%   End of Fig 6                                                   %%%
%%%%%%%%%%%%%%%%%%%%%%%%%%%%%%%%%%%%%%%%%%%%%%%%%%%%%%%%%%%%%%%%%%%%%%%%
%%%%%%%%%%%%%%%%%%%%%%%%%%%%%%%%%%%%%%%%%%%%%%%%%%%%%%%%%%%%%%%%%%%%%%%%
%%%   Fig 7: Performance versus allocation parameter                 %%%
%%%%%%%%%%%%%%%%%%%%%%%%%%%%%%%%%%%%%%%%%%%%%%%%%%%%%%%%%%%%%%%%%%%%%%%%
\begin{figure}[ht]
\begin{center}
\includegraphics[angle=0,width=0.8\textwidth]{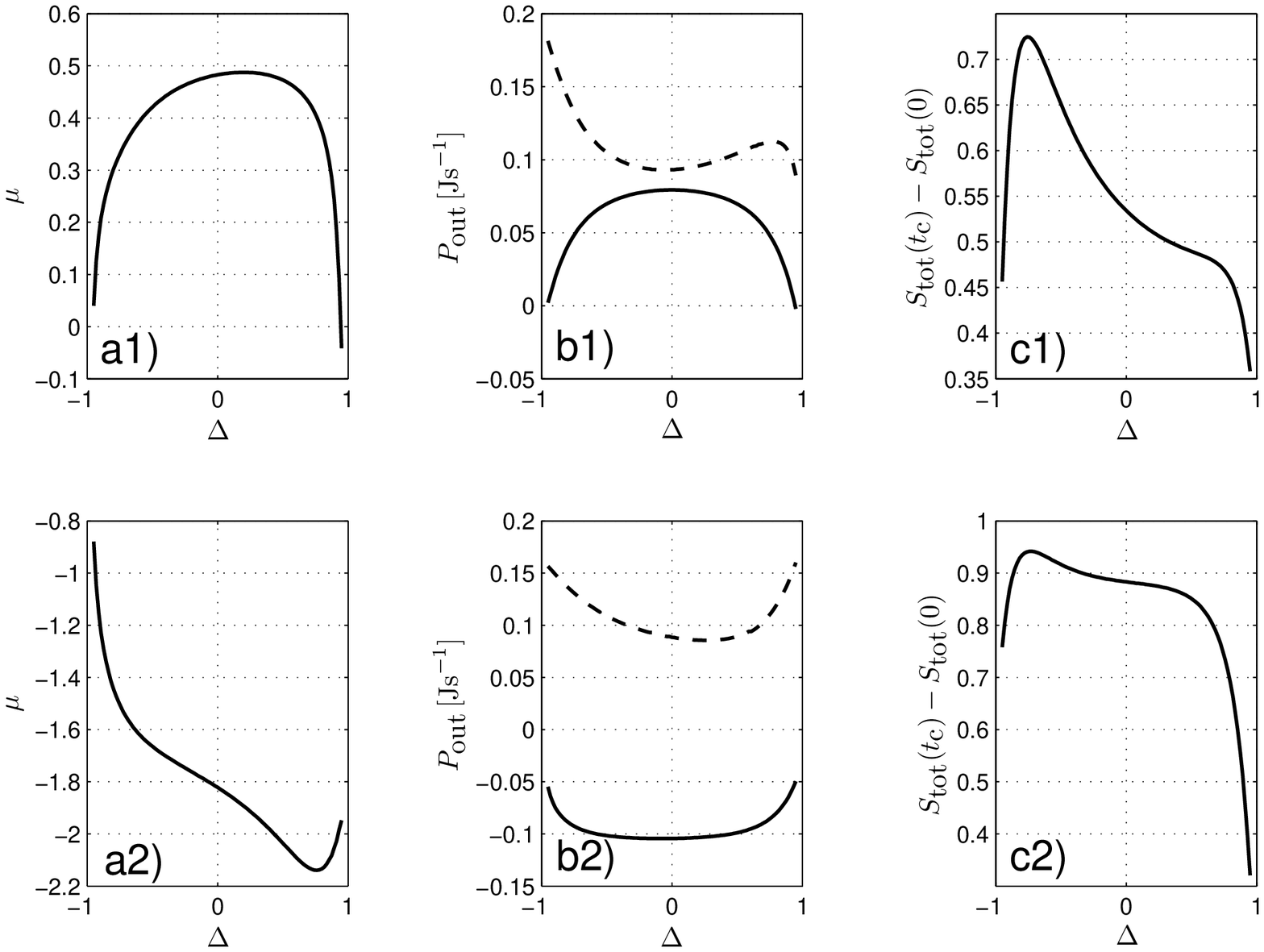}
\caption{ The engine performance characterized by the efficiency
  $\mu$, output power $\Pout$, and entropy production
  $S_{\mathrm{tot}}(t_{{\rm p}})-S_{\mathrm{tot}}(0)$, as a function
  of the asymmetry parameter $\Delta=(t_{+}-t_{-})/t_{{\rm p}}$ for a
  fixed period $t_{\rm p}=20\;\mathrm{s}$ and the same parameters
  $h_{1}=1\;\mathrm{J}$, $h_{2}=5\;\mathrm{J}$, $\nu = 1\;
  \mathrm{s}^{-1}$ as in the upper panel of figure~1. In a1)--c1) the
  bath during the first stroke is colder than during the second
  stroke: $\beta_{+}=0.5\;\mathrm{J}^{-1}$ and
  $\beta_{-}=0.1\;\mathrm{J}^{-1}$. Notice that the value $\Delta$ of
  maximum efficiency does not correspond to that of maximum output
  power. In a2)-c2) the reciprocal bath temperatures are interchanged
  compared to cases a1)-c1), $\beta_{+}=0.1\;\mathrm{J}^{-1}$ and
  $\beta_{-}=0.5\;\mathrm{J}^{-1}$. The dashed curves in b1) and b2)
  show the standard deviation of the output power calculated from
  $\rho_{\rm p}(w,\tper)$.
\label{Performance_asymetry}
}
\end{center}
\end{figure}
%%%%%%%%%%%%%%%%%%%%%%%%%%%%%%%%%%%%%%%%%%%%%%%%%%%%%%%%%%%%%%%%%%%%%%%%
%%%   End of Fig 7                                                   %%%
%%%%%%%%%%%%%%%%%%%%%%%%%%%%%%%%%%%%%%%%%%%%%%%%%%%%%%%%%%%%%%%%%%%%%%%%

\section{Discussion}
\label{sec:V} The overall properties of the engine critically depend
on the two dimensionless parameters
$a_{\pm}=\nu/(2\beta_{\pm}|v_{\pm}|)$. We call them
\emph{reversibility parameters}. For a given branch, say the first
one, the parameter $a_{+}$ represents the ratio of two characteristic
time scales. The first one, $1/\nu$, is given by the attempt rate of
the internal transitions \cite{comm:relax}. The second scale is
proportional to the reciprocal driving velocity. Contrary to the first
scale, the second one is fully under the external control. Moreover,
the reversibility parameter is proportional to the absolute
temperature of the heat bath.

Let us first consider the work probability density (\ref{rhocycle})
within the first stroke in the case $h_{2}>h_{1}$, c.f.\ figure.~2a).
In essence, $\rho_{\rm p}(w,t)$ is given by a linear combination of
the functions equations~(\ref{g11})-(\ref{g22}). It vanishes outside
the common support $[\,-v_{+}t,v_{+}t\,]$ which broadens linearly in
time. Besides the continuous part located within the support, the
diagonal elements $g_{ii}(w,0,t,0)$ display a singular part
represented by delta functions at the borders of the support. The
delta functions correspond to the paths with no transitions between
the states. Specifically, the weight of the delta function located at
$w=v_{+}t$ represents the probability that the system starts in the
first state and remains there up to time $t$. The weight corresponding
to the first level decreases with increasing time and vanishes for
$t\rightarrow\infty$. On the contrary, the weight of the delta
function at $-v_{+}t$ approaches the nonzero limit
$2\beta_{+}/(1+c_{+})^{a_{+}}$ for $t\rightarrow\infty$, which is the
probability that the a path starts in the second state \emph{and}
never leaves it.

Within the second stroke, the density $\rho_{\rm p}(w,t)$ results from
the integral of the propagators for the individual strokes, c.f.\
equation~(\ref{LimitCyclePropagatorG}). Due to the integration, the
singular parts of the cycle propagator $\matice{G}_+(w,0;t)$ are now
situated inside the support, at the values
$w=-v_{+}t_{+}+|v_{-}|(t-t_{+})$ and $w=v_{+}t_{+}-|v_{-}|(t-t_{+})$.
The two delta functions approach each other and, upon completing the
cycle, they coincide at the point $w = 0$. The nonsingular component
of the density is no more continuous \cite{comm:nonsignular}. The
jumps are located at the positions of the delta functions and their
magnitudes correspond to the weights of the delta functions (for a
discussion of the origin of these jumps, see \cite{Einax2009}).

If both reversibility parameters $a_{\pm}$ are small, the isothermal
processes during both branches strongly differs from the equilibrium
ones. The signature of this case is a flat continuous component of the
density $\rho_{\rm p}(w,t)$ and a well pronounced singular part. The
strongly irreversible dynamics occurs if one or more of the following
conditions hold. First, if $\nu$ is small, the transitions are rare
and the occupation probabilities of the individual energy levels are
effectively frozen during long periods of time. Therefore they lag
behind the Boltzmann distribution which would correspond to the
instantaneous positions of the energy levels. More precisely, the
population of the ascending (descending) energy level is larger
(smaller) than it would be during the corresponding reversible
process. As a result, the mean work done on the system is necessarily
larger than the equilibrium work. Secondly, a similar situation occurs
for large driving velocities $v_{\pm}$. Due to the rapid motion of the
energy levels, the occupation probabilities again lag behind the
equilibrium ones. Thirdly, the strong irreversibility occurs also in
the low temperature limit. In the limit $a_{\pm}\rightarrow 0$, the
continuous part vanishes and $\rho_{\rm p}(w,t_{\rm p})=\delta(w)$.

In the opposite case of large reversibility parameters $a_{\pm}$, both
branches in the $p\!-\!\!E$ plane are located close to the reversible
isotherms. The singular part of the density $\rho_{\rm p}(w,t)$ is
suppressed and the continuous part exhibits a well pronounced peak.
From general considerations \cite{Speck/Seifert:2004}, the density
must approach a Gaussian shape. Our results allow a detailed study of
this approach. Let us denote as $F(\beta,E)$ the free energy of a two
level system with energies $\pm E$ at temperature $T=1/(k_{{\rm
    B}}\beta$, i.e., $F(\beta,E)=-\frac{1}{\beta}\ln[2\cosh(\beta
E)]$. Let us further define
\begin{eqnarray}
\fl W_{\mathrm{rev}}(t)=\left\{ \begin{array}{ll}
F(\beta_{+},E_{1}(t))-F(\beta_{+},E_{1}(0)),& t\in[0,t_{+}], \nonumber\\
F(\beta_{-},E_{1}(t))-F(\beta_{-},E_{1}(t_{+}))+
F(\beta_{+},E_{1}(t_{+}))-F(\beta_{+},E_{1}(0)),
& t\in[t_{+},t_{{\rm p}}]. \nonumber
\end{array} \right. \\
\label{WmaxTime}
\end{eqnarray}
This is simply the reversible work done on the system if we transform
its state from the initial equilibrium state (with the energies fixed
at $\pm E_{1}(0)$) to another equilibrium state (with the energies
fixed at the values $\pm E_{1}(t)$). For large reversibility
parameters $a_{\pm}$, the peak of the work density $\rho_{\rm
  p}(w,t_{\rm p})$ occurs in the vicinity of the value
$W_{\mathrm{rev}}(t)$ and with increasing $a_{\pm}$, the peak
collapses to a delta function,
\begin{equation}
\label{rhoLimit}
\lim_{a_{\pm}\rightarrow \infty}\rho_{\rm p}(w,t)=\delta(w-W_{\mathrm{rev}}(t))\,\,.
\end{equation}

The main features of the heat probability density
$\chi_{\mathrm{p}}(q,t)$ from equation~(\ref{rhoHeatCycle}) are, as we
have seen in section~\ref{sec:III}, closely related to the work
through simple shifts of the independent variable $q$. However, there
are some interesting differences. While the work is conditioned by the
external driving, the heat exchange occurs as a consequence of the
transitions between the system states. The instantaneous positions of
the energies at the instant of the transition give the magnitude of
the heat exchange related with the given transition. From this
perspective, if there are no transitions, the exchanged heat is zero.
As a consequence, the singular part of the probability density
$\chi_{\mathrm{p}}(q,t)$ is always situated at $q=0$ and the weight of
the delta function at origin equals the sum of the weights of the
delta functions in the work density $\rho_{\rm p}(w,t)$. The support
of the heat density is given by the largest possible value of the
level splitting during the limit cycle. Within the first stroke the
support broadens linearly with time as
$[-2h_{1}-2v_{+}t,2h_{1}+2v_{+}t]$, up to its maximum width $[-2 h_2,
2 h_2]$ at the end of the stroke. Within the second stroke the energy
difference decreases and the support remains unchanged. The
non-singular part of the heat density always displays discontinuities
inside the support, even during the first stroke. In contrast to
$\langle\,i\,|\,{\mathbb G}_{\rm p}(w,t)|\,j\,\rangle$, the individual
elements $\langle\,i\,|\,{\mathbb K}_{\rm p}(q,t)|\,j\,\rangle$ in
equation~(\ref{rhoHeatCycle}) have different supports.

In the the strongly reversible regime each element
$\langle\,i\,|\,{\mathbb G}_{\rm p}(w,t)|\,j\,\rangle$ exhibits a
Gaussian shape situated at $W_{\mathrm{rev}}(t)$. The shift
transformation maps the Gaussian function onto four different
positions depending on the specific matrix element
$\langle\,i\,|\,{\mathbb K}_{\rm p}(q,t)|\,j\,\rangle$ in question. In
the reversible limit we have
\begin{equation}
\label{chilimit}
\lim_{a_{\pm}\rightarrow \infty}\langle\,i\,|\,{\mathbb K}_{\rm
  p}(q,t)|\,j\,
\rangle=
\delta\left(\,q-u_{ij}(t)+W_{\mathrm{rev}}(t)\,\right)\,\,.
\end{equation}
Using this form in equation~(\ref{rhoHeatCycle}) and calculating the
mean accepted heat, we get $Q(t)=U(t)-U(0)-W_{\mathrm{rev}}(t)$. In
the opposite limit, if $a_{\pm}\rightarrow 0$, we have
$\chi_{\mathrm{p}}(q,t)\rightarrow\delta(q)$ for any $t$.

According to the second law of thermodynamics, the mean work
$W(t)=\langle\,\mathsf{W}(t)\,\rangle$ must fulfill $|W(t)|\ge |W_{\rm
  rev}(t)|$. On the other hand, there always exists a fraction of the
paths which, individually, display the inequality $|\widetilde{w}({\rm
  path},t)|<|W_{\rm rev}(t)|$, where $\widetilde{w}({\rm path},t)$
denotes the work done on the system if it evolves along the indicated
path. Using the exact work probability density, we can calculate the
total weight of these trajectories. Specifically, in the case $W_{\rm
  rev}(t)>0$,
\begin{equation}
\label{ViolationWeight}
\prob{\,\RW{W}(t)<W_{\rm rev}(t)\,}=
\int_{-\infty}^{W_{\rm rev}(t)}\,\der{w}\,\rho_{\rm p}(w,t)\,\,.
\end{equation}
If $W_{\rm rev}(t)<0$, we would have to integrate over the interval
$(W_{\rm rev}(t),\infty,)$.

Let us finally note that in view of the rather complex structure of
the work and heat probability densities, we performed several
independent tests. First of all, the densities $\rho_{\rm p}(w,t)$ and
$\chi_{\mathrm{p}}(q,t)$ must be nonnegative functions fulfilling the
normalization conditions, e.g.\
$\int_{-\infty}^{\infty}\,\der{w}\,\rho_{\rm p}(w,t)=1$ for any
$t\in[0,\tper]$. Secondly, we have two different procedures to
calculate the first moment $W(t)=\langle\,\mathsf{W}(t)\,\rangle$. One
can either start with the density $\rho_{\rm p}(w,t)$ and evaluate the
required $w$-integral, or one directly employs the solution of the
rate equation as in Sec.\ IV. Another inspection is based on the
Jarzynski identity \cite{Jarzynski:1997,Crooks:1999}. In our setting,
consider the case $\beta_{\pm}=\beta$. After completing the cycle, the
system returns to the original state. Therefore we have
$W_{\mathrm{rev}}(t_{\rm p})=F(\beta,E_{1}(t_{\rm
  p}))-F(\beta,E_{1}(0))=0$ and the Jarzynski identity reduces to
$\langle\,\exp\left[-\beta\mathsf{W}(t_{\rm p})\right]\,\rangle=1$.
Using the explicit form of the work probability density we have
verified that the integral
$\int_{-\infty}^{\infty}\,\der{w}\,\exp(-\beta w)\rho_{\rm p}(w,t_{\rm
  p})$ actually equals one. Finally, we have studied the probability
densities $\rho_{\rm p}(w,t)$, $\chi_{\mathrm{p}}(q,t)$ by computer
simulation. In fact, we have developed two exact simulation methods.
Each of them uses a specific algorithm to generate paths of the
time-non-homogeneous Markov process $\mathsf{D}(t)$. Parts of these
simulation results have been published in \cite{Einax2009} and confirm
the analytical results.

\section{Conclusions}
\label{sec:VI} We have investigated a simple example of a microscopic
heat engine, which is exactly solvable. Based on mean thermodynamic
quantities, the engine performance is characterized by the occupation
probabilities of the energy levels following from the master equation.
The more challenging exact calculation of the work and heat
probability densities allowed us to study the fluctuation properties
in detail. A notable result is that the engine can be tuned to
maximize its output power, but the fluctuations of this quantity in
the corresponding optimal regime of control parameters are
comparatively high.

The present setting can be expanded in various directions. One can
address various problems concerning the thermodynamic optimization.
Another option would be the embodiment of additional (e.g., adiabatic)
branches. The role of the working medium can be assigned to other
systems that exhibit more complicated dynamics (e.g., diffusing
particles in the presence of time-dependent forces, or, variants of
the generalized master equation). It would be also interesting to
investigate settings with a nonlinear driving of the energy levels. A
nontrivial generalization would be the inclusion of a third energy
level. Having the three levels one can couple the system (different
pairs of forth-back transitions between the levels)
\emph{simultaneously} to reservoirs at different temperatures, so that
the system approaches a non-equilibrium steady state without driving
\cite{Zia2002}. Including a driving and forming an operational cycle,
there is no serious obstacle in repeating the present analysis for
this system, which has some additional intriguing properties compared
to the two-level system considered here (as, e.g., negative specific
heats).

Another possibility is an incorporation of specific forms of
transition rates \cite{Manosas:2009} that describe the stretching of
biomolecules in some realistic manner. In such problem, the histogram
of the work is experimentally accessible \cite{Manosas:2009}.
Particularly, in the experiments one can also determine the
probability of having certain number of transitions between the folded
and the unfolded conformation of the biomolecule during its mechanical
stretching \cite{Manosas:2009}. In our formulation, this information
is encoded in the counting statistics of the underlying random point
process \cite{Snyder1975} and can be extracted from the perturbation
expansion of the propagators which solve our dynamical equations.
Calculations in this direction are in progress and will be reported
elsewhere.

\ack Support of this work by the Ministry of Education of the Czech
Republic (project No. MSM 0021620835) and by the Grant Agency of the
Czech Republic (grant No. 202/07/0404) is gratefully acknowledged.


\begin{thebibliography}{99}
\bibitem{comm:relax} Note that due to the choice of the Glauber rates
  in equation~(\ref{transferrates}), the relaxation rate
  $[\lambda_1(t)+\lambda_2(t)]$ (for frozen energy levels at any time
  instant $t$) is bounded by $2\nu$.
\bibitem{comm:nonsignular} Only delta-functions are referred to as
  singularities here.
\end{thebibliography}
\end{document}